\documentclass[iop]{emulateapj}

\usepackage{graphicx}
\usepackage{epsfig}
\usepackage{epstopdf}
\usepackage[colorlinks=true,urlcolor=blue,citecolor=blue,linkcolor=blue]{hyperref}
\usepackage[T1]{fontenc}
\usepackage{amsmath}

\newcommand{\CaIIIR}{\ion{Ca}{2}~8542\AA}
\newcommand{\Halpha}{H\ensuremath{\alpha}}
\newcommand{\Mgk}{\ion{Mg}{2}~k}

\newcommand{\CaK}{\ion{Ca}{2}~K}
\newcommand{\CaKwvl}{\ion{Ca}{2}~K~3934\AA}
\newcommand{\SiIVs}{\ion{Si}{4}~1394\AA}
\newcommand{\SiIVl}{\ion{Si}{4}~1403\AA}
\newcommand{\kms}{km~s$^{-1}$}
\newcommand{\acknow}{IRIS is a NASA small explorer mission developed
  and operated by LMSAL with mission operations executed at NASA Ames
  Research center and major contributions to downlink communications
  funded by ESA and the Norwegian Space Centre.
The Swedish 1-m Solar Telescope is operated on the island of La Palma
by the Institute for Solar Physics (ISP) of Stockholm University in
the Spanish Observatorio del Roque de los Muchachos of the Instituto
de Astrof\' isica de Canarias.
This research was supported by the Research Council of Norway (project
number 250810) and through its Centres of Excellence scheme, project
number 262622. 
B.D.P. and J.M.S are supported by NASA contract NNG09FA40C (IRIS) and
grants NNH15ZDA001N-HSR and NNX16AG90G.
The numerical part of this work has been performed on computing
facilities from the programme for supercomputing of the Research
Council of Norway,
MareNostrum (BSC/CNS/RES, Spain), TeideHPC (ITER, Spain), and  
the Pleiades cluster through the computing project s1061 from the High 
End Computing (HEC) division of NASA. 
JdlCR is supported by grants from the Swedish Research Council
(2015-03994), the Swedish National Space Board (128/15) and the
Swedish Civil Contingencies Agency (MSB).  This research was supported
by the CHROMOBS grant of the Knut och Alice Wallenberg foundation.
D.N.S. is supported by the Spanish Ministry of Economy and
Competitiveness through projects AYA2011-24808 and AYA2014-55078-P.
SJ acknowledges support from the European Research Council (ERC) under
the European Union's Horizon 2020 research and innovation program
(grant agreement No. 682462).
This work profited from discussions at the meeting ``Solar UV bursts -
a new insight to magnetic reconnection" at the International Space
Science Institute (ISSI) in Bern.  }

\begin{document}

\title{Intermittent reconnection and plasmoids in UV bursts in the low solar
  atmosphere }

\author{L. Rouppe van der Voort\altaffilmark{1,2}}
\author{B.~De~Pontieu\altaffilmark{3,1,2}} 
\author{G.B.~Scharmer\altaffilmark{4,5}}
\author{J.~de~la~Cruz~Rodr{\'i}guez\altaffilmark{4}}
\author{J.~Mart{\'i}nez-Sykora\altaffilmark{3,6}}
\author{D.~N{\'o}brega-Siverio\altaffilmark{7,8}}
\author{L.J.~Guo\altaffilmark{3,6}}
\author{S.~Jafarzadeh\altaffilmark{1,2}} 
\author{T.M.D.~Pereira\altaffilmark{1,2}}
\author{V.H.~Hansteen\altaffilmark{1,2}}
\author{M.~Carlsson\altaffilmark{1,2}}
\author{G.~Vissers\altaffilmark{4}}

\affil{\altaffilmark{1}Rosseland Centre for Solar Physics,
  University of Oslo, %
  P.O. Box 1029 Blindern, NO-0315 Oslo, Norway}  
\affil{\altaffilmark{2}Institute of Theoretical Astrophysics,
  University of Oslo, %
  P.O. Box 1029 Blindern, NO-0315 Oslo, Norway}
\affil{\altaffilmark{3}Lockheed Martin Solar \& Astrophysics Lab, Org.\ A021S,
  Bldg.\ 252, 3251 Hanover Street Palo Alto, CA~94304 USA}
\affil{\altaffilmark{4}Institute for Solar Physics, Dept. of Astronomy, Stockholm University, 
  AlbaNova University Center, SE-10691, Stockholm, Sweden}
\affil{\altaffilmark{5}Royal Swedish Academy of Sciences, Box 50005, SE-10405, Stockholm, Sweden}
\affil{\altaffilmark{6}Bay Area Environmental Research Institute, Petaluma, CA 94952, USA}
\affil{\altaffilmark{7}Instituto de Astrof{\'i}sica de Canarias, Via Lactea, s/n, E-38205 La Laguna (Tenerife), Spain}
\affil{\altaffilmark{8}Department of Astrophysics, Universidad de La Laguna, E-38200 La Laguna (Tenerife), Spain}

\begin{abstract}
  Magnetic reconnection is thought to drive a wide variety of dynamic
  phenomena in the solar atmosphere. Yet the detailed physical
  mechanisms driving reconnection are difficult to discern in the
  remote sensing observations that are used to study the solar
  atmosphere.  In this paper we exploit the high-resolution
  instruments Interface Region Imaging Spectrograph (IRIS) and the new
  CHROMIS Fabry-P\'erot instrument at the Swedish 1-m Solar Telescope
  (SST) to identify the intermittency of magnetic reconnection and its
  association with the formation of plasmoids in so-called UV bursts
  in the low solar atmosphere. The \ion{Si}{4}~1403\AA\ UV burst
  spectra from the transition region show evidence of highly broadened
  line profiles with often non-Gaussian and triangular shapes, in
  addition to signatures of bidirectional flows. Such profiles had
  previously been linked, in idealized numerical simulations, to
  magnetic reconnection driven by the plasmoid
  instability. Simultaneous CHROMIS images in the chromospheric
  \CaKwvl\ 
  line now provide compelling evidence for the presence of plasmoids,
  by revealing highly dynamic and rapidly moving brightenings that are
  smaller than 0.2\arcsec\ and that evolve on timescales of order
  seconds.  Our interpretation of the observations is supported by
  detailed comparisons with synthetic observables from advanced
  numerical simulations of magnetic reconnection and associated
  plasmoids in the chromosphere. Our results highlight how
  subarcsecond imaging spectroscopy sensitive to a wide range of
  temperatures combined with advanced numerical simulations that are
  realistic enough to compare with observations can directly reveal
  the small-scale physical processes that drive the wide range of
  phenomena in the solar atmosphere.
\end{abstract} 




\section{Introduction}
\label{sec:intro}
Magnetic reconnection is thought to be the mechanism responsible for
releasing magnetic energy in a wide range of solar transient
phenomena, from flares and coronal mass ejections on the largest
scales, to jets, surges and UV bursts on spatial scales of just a few
arcseconds or smaller. It is clear from the observed temporal scales
and plasma conditions in this wide range of solar events that
classical reconnection, such as the Sweet-Parker mechanism, acts too
slowly (by orders of magnitude) to explain the rapid release of energy
that is observed \citep{Priest2014}. Recent theoretical work indicates
that the Sweet-Parker reconnection can turn into fast reconnection
through the tearing mode instability and formation of magnetic islands
or plasmoids \citep[e.g.][]{Loureiro2007,Bhattacharjee2009}.
Observational support for such plasmoids has been found from ``small''
bright blob-like features in imaging observations of CMEs and flares
\citep[e.g.][]{Ko2003,Lin2005,Lin2007,Lin2008,Milligan2010,Liu2013},
coronal jets \citep[e.g.][]{Zhang2014,Zhang2016,Zhang2017}, and
chromospheric anemone jets \citep{Singh2012}.  Whether the observed
bright blobs truly are magnetic islands as in the strict definition of
plasmoids, requires measurements of the magnetic field that are
presently not feasible.  Recently, there have also been suggestions
that the presence of plasmoids and fast reconnection can be deduced
from spectroscopic observations of transition region explosive events
which often show non-Gaussian and/or triangular-shaped profiles with
broad wings \citep{Innes2015}. Explosive events are highly dynamic
events, visible in transition region lines \citep{Dere1989}, that are
much smaller than the events described above, with typical total sizes
of only a few arcseconds.  They are thought to be driven by
reconnection
\citep{1997Natur.386..811I}. 
The presence of the plasmoid instability has been deduced from the
spectral line shapes since idealized numerical simulations suggest
that the piling up of plasmoids of a wide range of sizes naturally
leads to triangular or non-Gaussian line shapes with broad wings
around the current sheet, in addition to strong bidirectional flows in
the neighboring outflow regions \citep{Guo2017}.  If such line
profiles are indeed a signature of the plasmoid instability, the
\citet{Innes2015} results would significantly expand the diagnostic
capability of magnetic reconnection through remote sensing of the
solar atmosphere. First, spectroscopic measurements provide access to
velocities and densities, key constraints for theoretical models of
reconnection.  Second, and more importantly, this would allow the
identification of the plasmoid instability along the line-of-sight
without the need for identifying blob-like features in imaging
data. The latter is very difficult with most current imaging
instruments because the bulk of the plasmoids in such arcsecond-scale
events are expected to occur on much smaller spatial scales than can
be resolved. This is because the plasmoid instability is expected to
cascade down to very small spatial scales 
\citep[of order the ion inertial length,][]{Ni2015}.
However, blob-like features indicative of plasmoids have not been
directly seen in imaging data of such explosive events, because to
date imaging instruments have lacked the spatial resolution.

In this paper we focus on simultaneous spectroscopic and imaging
observations of so-called ``IRIS bombs''
\citep{Peter2014,Vissers2015b} or UV bursts. These events were
discovered with the Interface Region Imaging Spectrograph
\citep[IRIS,][]{De-Pontieu2014a} and are a subset of ``transition
region'' explosive events. The main difference with classical
explosive events is that they are formed much lower in the atmosphere
as evidenced by absorption from a \ion{Ni}{2} line in the blue wing of
\SiIVs. 
This absorption line is formed in cool,
chromospheric plasma (e.g., from overlying plasma in the magnetic
canopy) suggesting that the \ion{Si}{4} emission, traditionally viewed
as transition region diagnostic, originates from photospheric or low
chromospheric heights.
Comparisons with synthetic observables from advanced ``realistic'' 3D
numerical simulations show a remarkable correspondence with the
observations
\citep{2017ApJ...839...22H} 
and suggest that these events occur as a result of reconnection in the
low solar atmosphere when new magnetic flux emerges.

The main difference of our current work with these previous
observations and modeling efforts is the spatial resolution of the
imaging observations and the numerical resolution of the
simulation. Here we exploit the advent of the CHROMIS instrument, a
Fabry-P\'erot interferometer that operates at the Swedish 1-m Solar
Telescope
\citep[SST,][]{2003SPIE.4853..341S} 
in the blue part of the visible spectrum, allowing narrowband images
in the \CaKwvl\ 
spectral line with unprecedented spatial
resolution of 0\farcs08 
or 60 km. Such a resolution is high enough
to, for the first time, directly detect blob-like features in UV
bursts 
that also show triangular shaped \SiIVs\ profiles, 
thereby providing strong support for
the previous interpretation of plasmoid instability driven
reconnection. This is further confirmed by comparison with advanced
numerical simulations
revealing the presence of plasmoids in UV burst-like events.
Plasmoids have been seen previously in
several solar MHD simulations, including of chromospheric anemone jets
\citep[e.g.][]{Yang2013}, surges \citep{Nobrega-Siverio2016}, and
coronal jets \citep[e.g.][]{Ni2017}, but not for simulations that
focus on UV burst formation or that are realistic enough to allow
detailed comparisons with observables.
 
\section{Observations and Data Processing}
\label{sec:obs}

Active region NOAA AR12585 was observed in a coordinated campaign by
IRIS and SST on 2016 September 3, 4, and 5. Two Fabry-P\'erot tunable
filter instruments were employed at SST: CRISP
\citep{2008ApJ...689L..69S} 
on the red beam and CHROMIS, that saw first light during this
campaign, on the blue beam.
A detailed description of the CHROMIS instrument will be published in
a forthcoming paper by Scharmer and collaborators.
CHROMIS sampled two spectral lines: \CaKwvl\ and
H\ensuremath{\beta}.  For this study, we only focus on \CaK, 
which was sampled at 21 wavelength positions, within the range of
Doppler offsets of $\pm101$~\kms\ from line center and with
6~\kms\ steps (or 78~m\AA) between $\pm54$~\kms.  In addition, a
continuum position was sampled at 4000~\AA.  The temporal cadence was
13, 25, and 12~s for the three observing days, and the time spent to
sample \CaK\ 
was 8.1, 15.5, and 7.4~s respectively.
The CHROMIS transmission width is estimated to be 130~m\AA\ at
\CaK, the image scale is 0\farcs0376, and the diffraction
limit $\lambda/D$ is 0\farcs08 at 3934~\AA.
CRISP sampled H\ensuremath{\alpha} at 15 line positions between
$\pm$1.5~\AA,
and \CaIIIR\ 
at 21 line positions between
$\pm$1.75~\AA\ in spectropolarimetric mode.
The temporal cadence was 20~s.  On 2016 September 5, CRISP also
sampled the two \ion{Fe}{1} lines at 6301 and 6302~\AA, at 16 line
positions in spectropolarimetric mode, this decreased the temporal
cadence to 32~s.
We applied Multi-Object Multi-Frame Blind Deconvolution
\citep[MOMFBD,][]{2005SoPh..228..191V} 
image restoration separately to each individual spectral line scan,
which allowed, with the aid of the SST high order adaptive optics
system, for diffraction limited imaging during the best seeing
conditions.
CHROMIS includes a pair of cameras in phase-diversity mode for the
wideband channel (filter FWHM 13.2~\AA).
We followed the CRISPRED data reduction pipeline
\citep{2015A&A...573A..40D} which was adapted and extended for the
CHROMIS data
\citep[CHROMISRED, ][]{loefdahl17}.  

IRIS was running a so-called ``medium dense 16-step raster''
(observing program OBS-ID 3625503135) which has the spectrograph slit
cover an area of 5\farcs3$\times$60\arcsec\ with 16 continuous
0\farcs33 steps and a temporal cadence of 20~s.  The exposure time was
0.5~s, the spatial image scale 0\farcs166 pixel$^{-1}$, and the
spectral sampling 11~\kms\ for \ion{Si}{4} (4$\times$ binning). 
Slit-jaw images were recorded
in the 1400\AA\ (dominated by \ion{Si}{4}), 1330\AA\ (dominated by
\ion{C}{2}), and 2976\AA\ (\Mgk) channels at 10~s temporal
cadence.

The overlapping and aligned SST + IRIS data sets have durations of 27
min (September 3, starting 09:36:50 UT), 34 min (September 4, starting
08:19:17), and 30 min (September 5, starting 09:15:16).

\section{Numerical Simulations}
\label{sec:sim}

To interpret the observations we use a 2.5D radiative MHD
simulation using the Bifrost code
\citep{2011A&A...531A.154G} 
which includes thermal conduction along the magnetic field and
radiative losses from the photosphere to the corona
\citep{Hayek:2010ac,Carlsson:2012uq}.  The model spans from the upper
layers of the convection zone ($z=-2.6$~Mm, with the photosphere at
$z=0$~Mm) to the corona ($z=32$~Mm). The ambient magnetic field is
vertical, unipolar, and of strength 10~G. In addition, a twisted
magnetic flux tube ($6.3\cdot10^{18}$~Mx) is injected from the bottom
boundary which results in flux emergence in the simulated atmosphere
and the formation of a reconnection site between the emerged plasma
and the ambient field (see Fig.~\ref{fig:sim}). A detailed description
of this simulation is given in \citet{Nobrega-Siverio:2017sim}.

We calculate synthetic spectral profiles of \ion{Si}{4} assuming
non-equilibrium ionization \citep{Olluri13} and optically thin
conditions.  The upper atmosphere in the simulation has relatively low
density and the chromosphere, transition region and corona combined
has a \ion{Si}{4} line core optical thickness $\tau<0.1$ so that the
assumption of optically thin line formation is justified.  In
addition, synthetic \ion{Ca}{2} spectral profiles are calculated using
the RH code \citep{Uitenbroek:2001dq,Pereira:2015th} assuming a five
plus continuum level \ion{Ca}{2} atom in 1D, non-LTE and partial
redistribution
\citep{2012A&A...543A.109L}. 
The synthetic \ion{Si}{4} and \ion{Ca}{2} profiles have been degraded
both spatially and spectrally to match the observations.

\section{Results}
\label{sec:results}

\begin{figure*}[!t]
\begin{center}
\includegraphics[width=\textwidth]{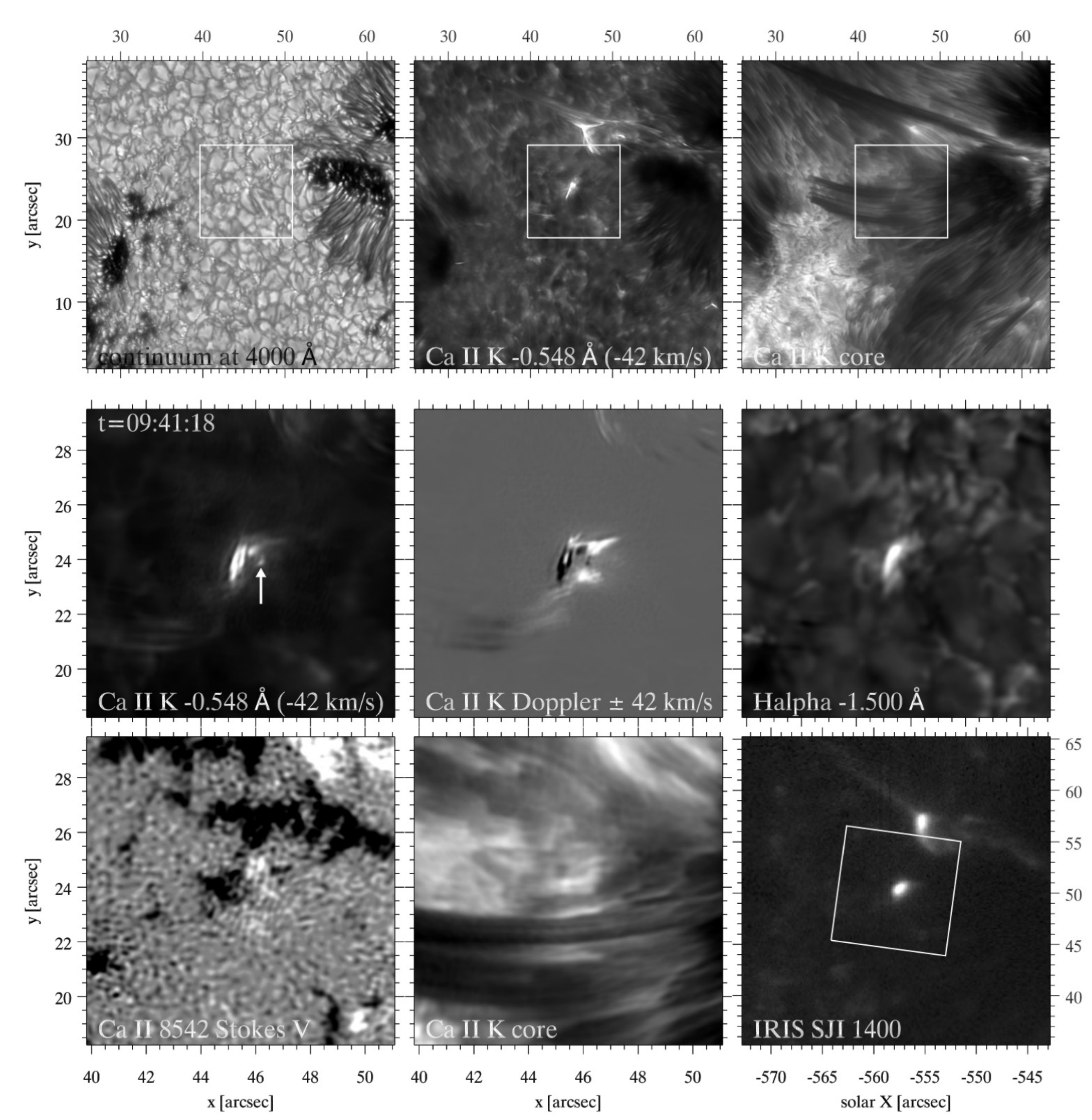}
\caption{Different imaging diagnostics for the 2016 September 3
  dataset. The top row shows large field-of-view (FOV) overview
  images.  The white box in the top row and the IRIS SJI 1400 image in
  the lower right marks the area centered on the UV burst shown in
  more detail in the SST images in the bottom rows.  All images are
  shown with coordinates of the CHROMIS FOV, except the IRIS SJI 1400
  image, which has heliocentric-cartesian coordinates.  The white
  arrow marks an isolated plasmoid-like blob. An animated version of
  the bottom two rows is available as online material
  at \url{http://folk.uio.no/rouppe/plasmoids_chromis/}.}
\label{fig:overview1}
\end{center}
\end{figure*}

\begin{figure*}[!t]
\begin{center}
\includegraphics[width=\textwidth]{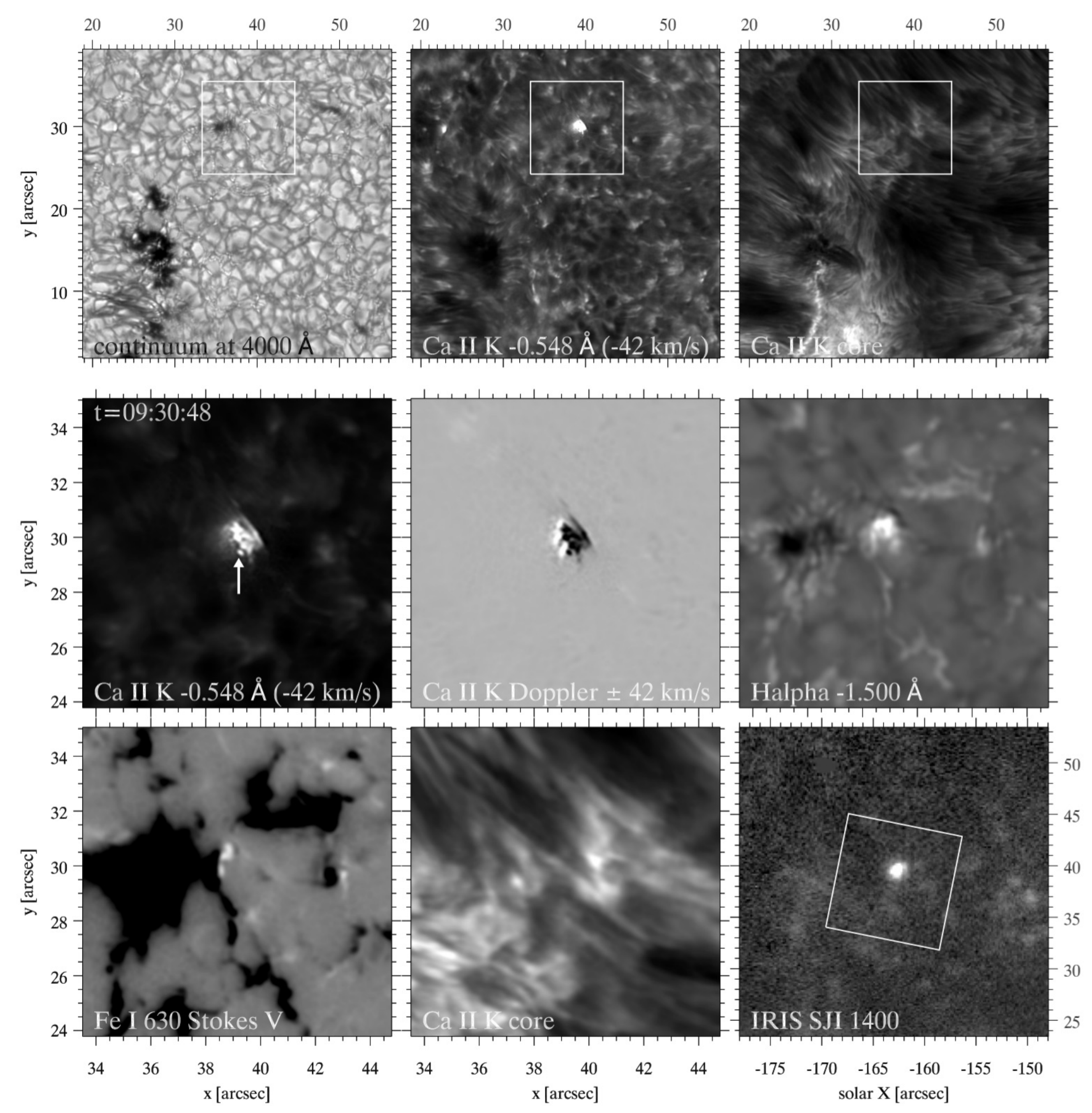}
\caption{As Fig.~\ref{fig:overview1} but for the 2016 September 5
  dataset. An animated version of the bottom two rows is available as
  online material at \url{http://folk.uio.no/rouppe/plasmoids_chromis/}.}
\label{fig:overview2}
\end{center}
\end{figure*}

\begin{figure*}[!t]
\begin{center}
\includegraphics[width=\textwidth]{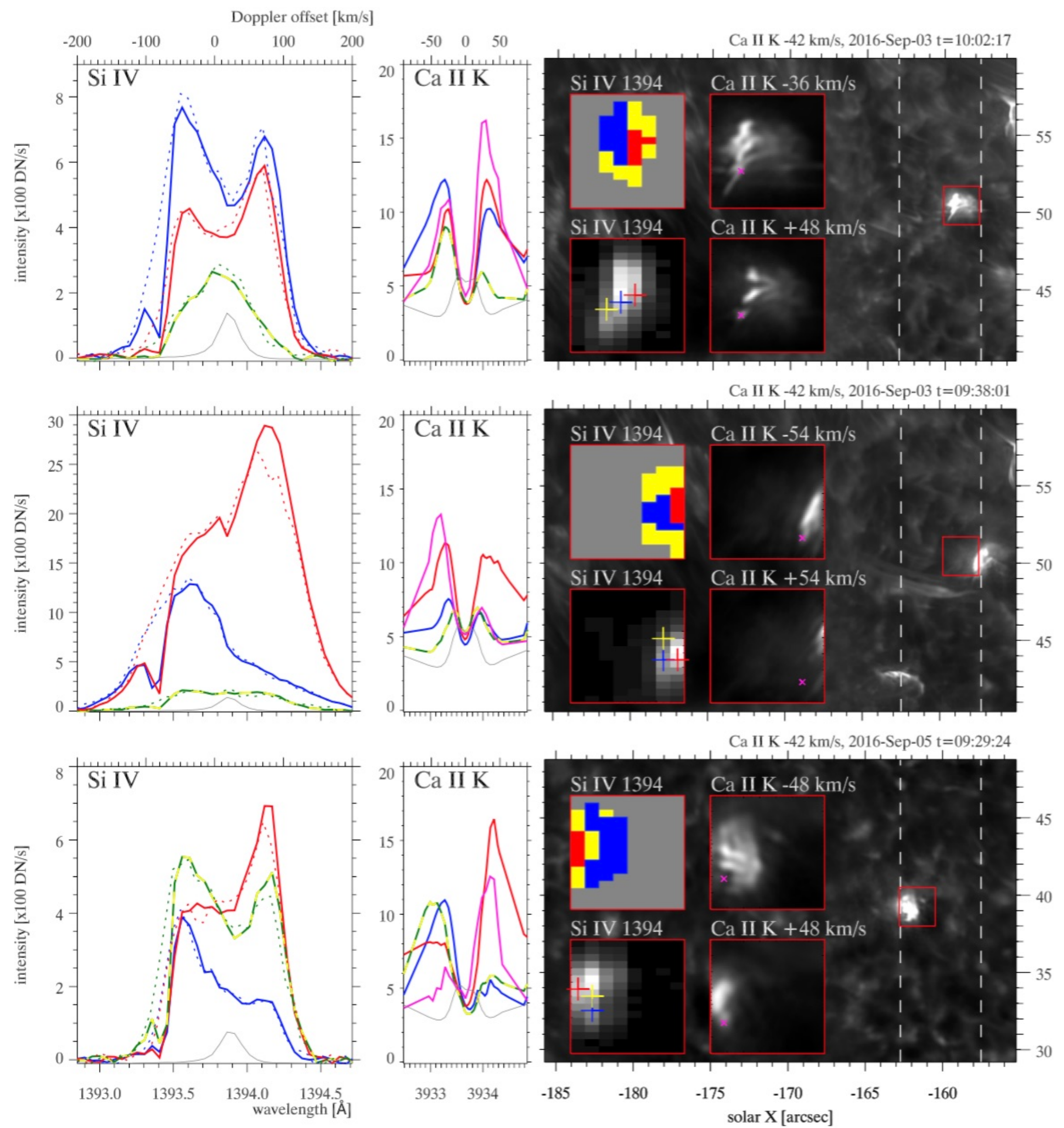}
\caption{UV burst spectral profiles and diagnostics. The left two
  panels show spectral profiles at 3 different spatial locations
  marked with 3 different colors: blue, red and green/yellow, all
  recorded during the same IRIS raster time. The 3 different spatial
  locations are in close proximity of each other, their locations are
  marked with crosses in the inset \ion{Si}{4} raster maps in the
  right panels. The solid lines in the left panels are for \SiIVs, the
  dotted lines are \SiIVl\ profiles, shifted in wavelength and
  multiplied with a factor 2 to compensate for the difference in the
  atomic transition's oscillator strength. The right panels show \CaK\
  profiles in arbitrary units and averaged over the IRIS pixel. The
  pink profile is from a CHROMIS pixel in a plasmoid marked with a
  pink cross in the \CaK\ insets in the right panels. The thin grey
  profiles are average spectral profiles for reference, the average
  \SiIVs\ profile is multiplied with a factor 10 for clarity. The wide
  right panel is a \CaK\ wing reference image, the vertical dashed
  lines mark the IRIS raster extent. The red box outlines the region
  for which inset images are shown at higher magnification. The lower
  left \SiIVs\ map is integrated intensity over the full spectral
  window, in the top left map colors mark \ion{Si}{4} profile
  classification: red/blue pixels have strong ($>30$~\kms) red/blue
  shifted components, yellow have broad or triangular non-gaussian
  profiles.}
\label{fig:sprast}
\end{center}
\end{figure*}

Visual inspection of the \CaK\ movies obtained with CHROMIS on
all three days reveal a bewildering variety of fine-scale, highly
dynamic phenomena, some of which evolve so rapidly that they are not
fully captured by the cadence (12 to 25~s) of the timeseries. This is
illustrated by the online animations that accompany
Figs.~\ref{fig:overview1} and \ref{fig:overview2}. These figures
highlight the two events that are the focus of this paper. These are
typical cases of UV bursts with strong emission in the wings of
H$\alpha$ and the \CaK\ line, as well as the IRIS SJI
1400\AA\ channel which is dominated by transition region \ion{Si}{4}
emission.  The Stokes V magnetogram movies show that in both cases the
UV bursts are caused by a sequence of events that starts with
emergence of flux (seen as an expanding bubble in the magnetograms)
followed by cancellation of the newly emerged flux with the strong
pre-existing neighboring plage fields, most plausibly accompanied by
magnetic reconnection. The whole process occurs on timescales of order
5-15 min and is set by the evolution of the photospheric magnetic
field. The chromospheric and transition region response is much more
dynamic, as illustrated with the \CaK\ images from CHROMIS and
the SJI movies from IRIS. Both events appear to occur underneath a set
of fibrils that overlie the cancellation site as can be seen in the
\CaK\ core images. 
The response to the reconnection, i.e., the
UV burst, becomes visible in the wings of the chromospheric lines
\citep[as previously reported, e.g., by][]{Vissers2015b}.
In \Halpha\ this is manifested as typical Ellerman bomb profiles with
strongly enhanced wings.
The extremely high resolution enabled by CHROMIS allows us to now
better spatially resolve some of the substructure of these UV
bursts. We are struck by the spatial and temporal intermittency of the
brightenings in both the blue and red wings of \CaK. This
intermittency manifests itself in the wing images as tiny blob-like
features that rapidly change in position and sometimes in shape from
one timestep to the next, suggesting that the dynamic evolution is not
always properly tracked. Nevertheless we see several occasions when
blob-like features last for several timesteps and rapidly move away
from the cancellation site. These features are visible in the wings at
velocities of order 40~\kms\ as illustrated by the \CaK\
Dopplergrams in Figs.~\ref{fig:overview1} 
and \ref{fig:overview2} and
the spectral line profiles in Fig.~\ref{fig:sprast}. 
A white arrow marks an example of an isolated, FWHM 0\farcs14 wide
blob in Fig.~\ref{fig:overview1}, and a 0\farcs12 wide blob in
Fig.~\ref{fig:overview2}.
Both the line-of-sight (LOS) and plane-of-the-sky velocities are in
excess of the typical values for the speed of sound in the
chromosphere ($\sim15$~\kms), and of order the Alfv\'en speed,
suggesting a magnetic origin for the driving mechanism that
accelerates the plasma.

The blobs often disappear from the chromospheric passband within one
timestep, which, given the clear counterpart in the IRIS transition
region SJI channel, suggests that significant heating is occurring and
depletion of \ion{Ca}{2} due to ionization at high temperature.  The
transition region spectra associated with these events show profiles
that are common in classical explosive events \citep{Innes2015}. We
see profiles that are strongly Doppler shifted to the blue or red
(Fig.~\ref{fig:sprast} 
), as well as profiles that are more centered
around 0~\kms\ with very broad wings up to 100
\kms. These profiles are non-Gaussian and often show
triangular-shaped profiles, which \citet{Innes2015} and
\citet{Guo2017} have suggested are caused by reconnection driven by
the plasmoid instability.
The spatial pattern of these strongly broadened explosive event type
profiles in \ion{Si}{4} is illustrated in Fig.~\ref{fig:sprast} which
shows a multitude of non-Gaussian, flat-topped or triangular-shaped,
profiles, sometimes centered around 0~\kms\ (yellow), as well as
strongly blue-shifted and red-shifted profiles in the vicinity. All of
these observations are compatible with a scenario in which these
events are caused by plasmoid-mediated magnetic reconnection, likely
at a wide range of locations within the UV burst(s), with yellow
profiles caused by pile-up of a variety of plasmoids, and
red/blue-shifted profiles in the outflow regions
\citep{Innes2015,Guo2017}.
This tentative scenario is based on idealized numerical simulations,
but finds strong support from our discovery of the highly intermittent
nature of the UV burst brightenings mediated by dynamic, fine-scale
blob-like structures in the CHROMIS data.
 
\begin{figure*}[!t]
\begin{center}
\includegraphics[width=\textwidth]{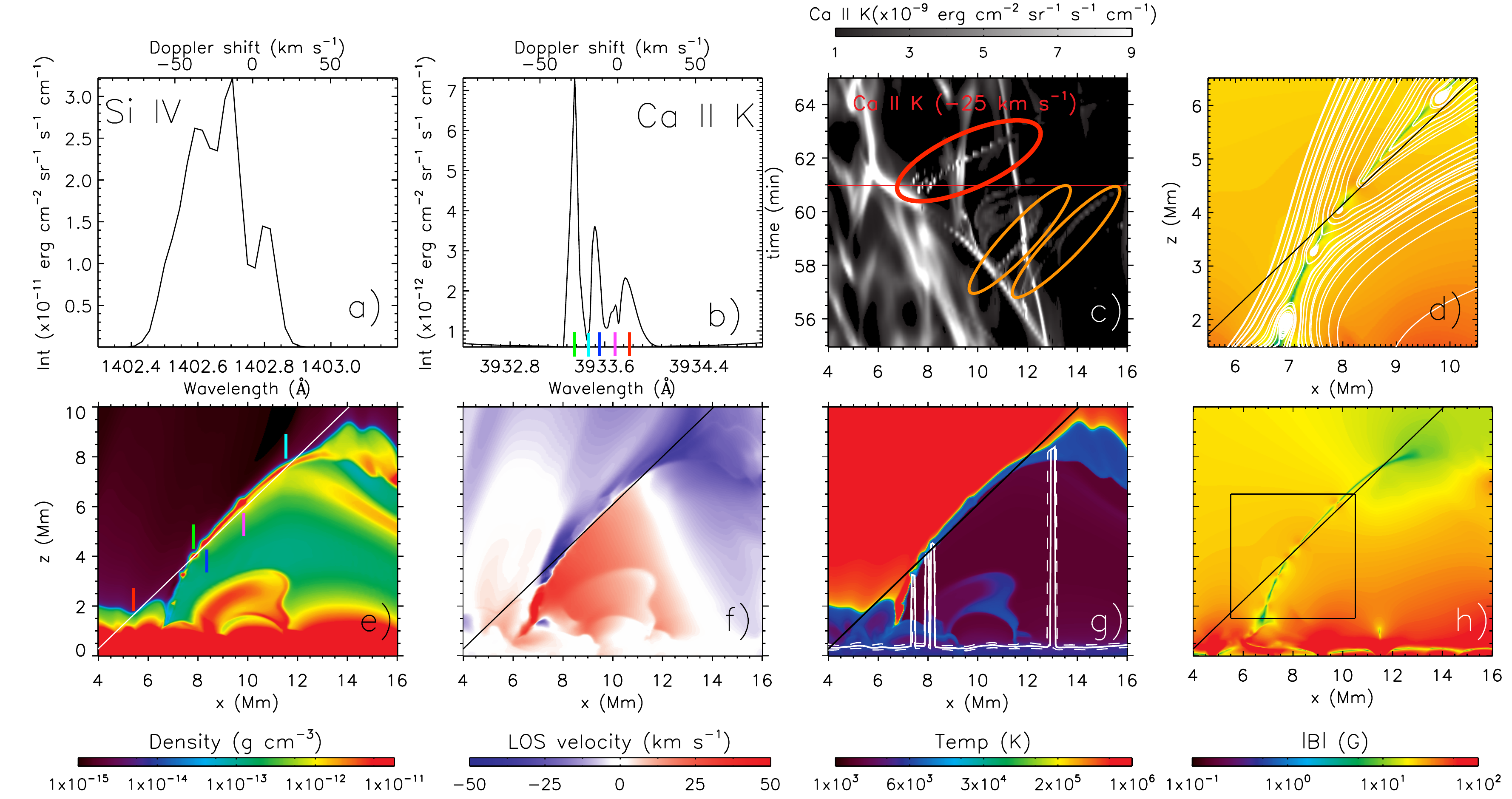}
\caption{Results from the Bifrost simulation. Synthetic spectral
  profiles of \ion{Si}{4} (a) and \ion{Ca}{2} (b) computed along the
  inclined line shown in panels (d)--(h) which crosses some plasmoids.
  Panel (c) shows the space-time plot of \ion{Ca}{2} blue wing
  ($25$~\kms) with the LOS along the vertical axis. The red horizontal
  line is marking the time of the other panels.  Maps of the density,
  velocity along the inclined LOS, and temperature are shown in panels
  (e), (f) and (g).  The white contours in panel (g) mark the heights
  of $\tau=0.3, 1, 3$ along the vertical LOS for the $-25$~\kms\ blue
  wing of \CaK.  Panels (d) and (h) show the absolute magnitude of the
  magnetic field, with (d) zooming in on plasmoids which can be
  recognized as magnetic islands by selected field lines (white
  contours).  The plasmoids are readily seen in the density map and
  correspond to very narrow \ion{Ca}{2} brightenings that travel from
  left to right in panel (c) and are high-lighted by ellipses: the red
  ellipse marks the plasmoids that produced the \CaK\ profile in panel
  (b), the orange ellipses mark earlier episodes with plasmoids
  visible in the \CaK\ blue wing.  Colored markers in panel (e)
  indicate the location of $\tau=1$ of the corresponding \CaK\
  spectral features in panel (b), the peaks at $-29, -16$, and
  $-3$~\kms\ (green, dark blue, and pink markers) are caused by
  different plasmoids.  An animated version of panels (c), (e), (f),
  and (g) is available as online material
  at \url{http://folk.uio.no/rouppe/plasmoids_chromis/}.}
\label{fig:sim}
\end{center}
\end{figure*}

Additional support for this scenario comes from advanced numerical
simulations in which flux emergence leads to magnetic reconnection
that results in a UV burst that is accompanied by a surge 
\citep[for more details, see][]{Nobrega-Siverio:2017sim}.
We focus here on the presence of plasmoids in this simulation. 
As shown in the online animation that accompanies Fig.~\ref{fig:sim},
the reconnection quickly develops a sequence of plasmoids that
propagate both upward and downward. Synthesis of \CaK\ and \ion{Si}{4}
observables from these simulations (Fig.~\ref{fig:sim}) shows notable
similarities with the observations.
We see emergent \CaK\ spectral line profiles that show enhanced
components in the wings, in a similar fashion to those shown in
Fig.~\ref{fig:sprast}: both in terms of relative enhancement and range
of Doppler shifts
(note, however, that
the synthetic profiles do not show a dark absorption core K3 like in
the observations since the simulation does not have dense overlying
fibrils).  The synthetic profiles are caused by reconnection outflows
filled with plasmoids which show significant redshifts and blueshifts.
The marking of the $\tau=1$ height in Fig.~\ref{fig:sim}g shows that
the plasmoids are responsible for the bright blobs in the \CaK\ wing.
The magnetic reconnection in the simulation leads to substantial
heating of plasma to at least transition region temperatures.
This results in strong \ion{Si}{4} brightenings.  The spectral line
profile that emerges from a LOS along the current sheet shows
substantially broadened profiles that often have non-Gaussian and
sometimes triangular shapes. These profiles show a good resemblance
with those observed (Fig.~\ref{fig:sprast}). Comparison with the
velocity panel in Fig.~\ref{fig:sim} shows that the broadened
\ion{Si}{4} comes about because the LOS includes multiple plasmoids
with the upward and downward moving plasmoids in the outflow regions
and the stationary plasmoids close to the current sheet. This confirms
the scenario proposed by \citet{Guo2017} using idealized simulations.

The synthetic \ion{Si}{4} profiles are narrower than the observed
ones, likely because the numerical resolution of the current
simulation is not high enough to capture the multitude of plasmoids
that in more idealized simulations \citep{Innes2015} with much higher
resolution lead to bright and broader spectral profiles.
This is supported by numerical experiments in which the resolution of
the current simulation is increased by a factor of two. Such
high-resolution runs show a significant increase of the number of
plasmoids, as expected from theoretical considerations, and more
triangular shaped broadened profiles.

\section{Conclusions}
\label{sec:conclusions}

We have found evidence for intermittent magnetic reconnection driven
by the plasmoid instability in UV bursts in the low solar
atmosphere. We exploit the extremely high spatial resolution of the
new CHROMIS instrument at the SST to reveal the presence of rapidly
evolving blob-like features at Alfv\'enic speeds in the \CaK\ line. We
see evidence for substantial heating to transition region temperatures
leading to highly broadened, non-Gaussian \ion{Si}{4} profiles
observed with IRIS. Such profiles had previously been associated with
reconnection mediated by the plasmoid instability based on idealized
numerical simulations \citep{Innes2015}. This has now been placed on a
solid footing by the direct observational evidence of tiny blob-like
features, and by advanced numerical simulations of UV bursts which
show remarkable similarities to the observations.  Our results
establish the presence of bright, non-Gaussian/triangular shaped
spectral line profiles in spectroscopic data as good proxies for
plasmoid-mediated reconnection. This finding will allow us to exploit
spectroscopy to diagnose small-scale plasma physical processes like
the plasmoid instability on spatial scales where imaging cannot reveal
whether reconnection occurs and/or determine the dominant mode of
reconnection.

\acknowledgements
\acknow


\end{document}